\documentclass[aps,preprint]{revtex4}%
\usepackage{amsfonts}
\usepackage{amsmath}
\usepackage{amssymb}
\usepackage{graphicx}%
\begin{document}
\title{New Non-Abelian Black Hole Solutions in Born-Infeld Gravity}
\author{S. Habib Mazharimousavi$^{\ast}$}
\author{M. Halilsoy$^{\dag}$}
\author{Z. Amirabi$^{\ddag}$}
\affiliation{Department of Physics, Eastern Mediterranean University,}
\affiliation{G. Magusa, north Cyprus, Mersin-10, Turkey}
\affiliation{$^{\ast}$habib.mazhari@emu.edu.tr}
\affiliation{$^{\dagger}$mustafa.halilsoy@emu.edu.tr}
\affiliation{$^{\ddagger}$zahra.amirabi@emu.edu.tr}

\begin{abstract}
We introduce new black hole solutions to the Einstein-Yang-Mills-Born-Infeld
(EYMBI), Einstein-Yang-Mills-Born-Infeld-Gauss-Bonnet (EYMBIGB) and
Einstein-Yang-Mills-Born-Infeld-Gauss-Bonnet-Lovelock (EYMBIGBL) gravities in
higher dimensions $N\geq5$ to investigate the roles of Born-Infeld parameter
$\beta$. It is shown that, these solutions in the limits of $\beta
\rightarrow0,$ and $\beta\rightarrow\infty,$ represent pure gravity and
gravity coupled with Yang-Mills fields, respectively. For $0<\beta<\infty$ it
yields a variety of black holes, supporting even regular ones at $r=0$.

\end{abstract}
\maketitle

\section{Introduction}

Historically, Born-Infeld (BI) nonlinear electrodynamics model was formulated
in 1934\cite{1}. Since it has been proposed as a viable model in low energy
string theory, BI electrodynamics has attracted much attention from the fronts
of both string theory and cosmology\cite{2}. While classical electrodynamics
due to Maxwell is a linear theory obeying the principle of superposition, in
the latter these properties are not valid any more. In this respect the BI
electrodynamics is comparable with the other nonlinear theories of physics
such as Yang-Mills (YM) and gravitation. Finding plane wave solutions in such
a theory , for instance, in the presence of boundaries and/or background
effects becomes a difficult task. Coupling of BI electrodynamics to gravity
has given birth to a new theory known as the Einstein-Born-Infeld (EBI)
gravity which found applications in string theory. The BI version of
electromagnetism is already in the form of a string Lagrangian, i.e. square
root of a determinant, living inside higher dimensional worlds of branes.
Addition of Higgs field and investigating its monopole solutions become
equally attractive for the field theorists in the realm of confinement related
problems\cite{3}. We recall that the original BI electrodynamics was
introduced in order to resolve the self-energy divergence in the Coulomb
problem. With the advent of quantum electrodynamics this feature of BI theory
was almost forgotten. Coincidentally, beside other things, string theory was
also introduced to eliminate divergences due to point-like structures.
Combination of these two theories (i.e. BI and string theory) is expected
naturally to yield finite physical results. BI action in supergravity admits
solitonic solutions known as D-branes which form the end points for open
strings. In this paper, however, we shall not address ourselves to D-branes or
dilatons, postponing these to a future study. A different theory, which will
establish our strategy in this paper, is to consider the EBI action in which
instead of the electromagnetic field we employ the non-Abelian YM field
\cite{4}. For this purpose we make use of a YM ansatz in the spherically
symmetric spacetime. Recently we have obtained such EYM black hole solutions
and extended it to the higher dimensional Gauss-Bonnet (GB) and Lovelock
theories \cite{5}. Our method of solving the YM equations was to generalize
the original Wu-Yang ansatz in N=4, to higher dimensions $(N\geq5).$ In this
ansatz the YM field is of magnetic type so that the invariant $F_{\lambda
\sigma}^{\left(  a\right)  \star}F^{\left(  a\right)  \lambda\sigma}=0,$ in
the action, leaving behind the term $F_{\lambda\sigma}^{\left(  a\right)
}F^{\left(  a\right)  \lambda\sigma}\neq0.$ As expected, employing YM instead
of the Maxwell field, accumulates different types of nonlinearities to yield,
altogether a highly nonlinear model of gravity a la' BI formalism. In the
proper limit $\beta\rightarrow\infty$, where $\beta$ is called the BI
parameter, we recover the Einstein-Hilbert action coupled with YM field in the
standard way. Einstein- Hilbert action constitutes the simplest geometrical
theory which involves mass as its parameter. Its geometrical/ topological
extensions employ higher order invariants with more parameters that provide
extra degrees of freedom in the theory. The Lovelock Lagrangian is the most
general Lagrangian that admits second order equations without invoking ghost
structures. By taking appropriate limits we recover all interesting cases
obtained so far. It is remarkable that three highly non-linear theories, such
as BI, YM and Lovelock gravity are brought together in a common Lagrangian
which admits exact solutions.

In this paper we address to the issue of black hole solutions in the EBI
action by incorporating YM fields in higher dimensions. Naturally the BI
parameter $\beta$ modifies the black holes and their thermodynamics
properties. Next, we consider the Gauss-Bonnet (GB) extension and search for
new features brought in by the topological properties of the GB theory. The
latter has the property that in the absence of a true cosmological constant
$\Lambda$, asymptotically it produces an effective one, $\Lambda_{eff}$ to
imitate the real one. In other words, the de-Sitter (dS) and Anti de-Sitter
(AdS) spacetimes which are of utmost importance in the conformal field theory
correspondence arise simply as boundary conditions of the spacetime. Inclusion
of the $\beta$ parameter adds further degrees of freedom to the theory. We
find, for example, that $\beta$ can be employed to construct / regulate black
hole horizons at wish. Extension to the third order Lovelock gravity, however,
restricts our exact solution such that in the absence of a real cosmological
constant it does not admit an effective one.

The paper is organized as follows: In Sec. $II$ we introduce the
Einstein-Yang-Mills-Born-Infeld (EYMBI) action, metric, YM ansatzes and the
resulting field equations. In the same section we find exact solutions of the
field equations in $N\geqslant5$. Sec. $III$ follows by introducing the
action, field equations and solutions for the $N\geqslant5$ dimensional
EYMBIGB theory. In Sec. $IV$ we follow the same patterns for the EYMBIGBL, in
which the abbreviation L refers to the third order Lovelock gravity. The paper
ends with concluding remarks in Sec. $V$.

\section{Field Equations and the metric ansatz for EYMBI gravity}

The $N\left(  =n+1\right)  -$dimensional action for
Einstein-Yang-Mills-Born-Infeld gravity with a cosmological constant $\Lambda$
is given by
\begin{equation}
S=\frac{1}{16\pi}\int_{\mathcal{M}}d^{n+1}x\sqrt{-g}\left(  R-\frac{n\left(
n-1\right)  }{3}\Lambda+L\left(  \mathbf{F}\right)  \right)  +\frac{1}{8\pi
}\int_{\partial\mathcal{M}}d^{n}x\sqrt{-\gamma}K\left(  \gamma\right)  ,
\end{equation}
in which the YMBI Lagrangian $L\left(  \mathbf{F}\right)  $ is given by%
\begin{equation}
L\left(  \mathbf{F}\right)  =4\beta^{2}\left(  1-\sqrt{1+\frac{\mathbf{Tr}%
(F_{\lambda\sigma}^{\left(  a\right)  }F^{\left(  a\right)  \lambda\sigma}%
)}{2\beta^{2}}+\frac{\mathbf{Tr}(F_{\lambda\sigma}^{\left(  a\right)  \star
}F^{\left(  a\right)  \lambda\sigma})^{2}}{16\beta^{4}}}\right)  ,
\end{equation}
where
\begin{equation}
\mathbf{Tr}(.)=\overset{n(n-1)/2}{\underset{a=1}{%
{\textstyle\sum}
}\left(  .\right)  }.
\end{equation}
Herein we are interested in the magnetically charged YM Ansatz in which
$\mathbf{Tr}(F_{\lambda\sigma}^{\left(  a\right)  \star}F^{\left(  a\right)
\lambda\sigma})$ $=0$ and therefore $L\left(  \mathbf{F}\right)  $ reduces to
the form%
\begin{equation}
L\left(  \mathbf{F}\right)  =4\beta^{2}\left(  1-\sqrt{1+\frac{\mathbf{Tr}%
(F_{\lambda\sigma}^{\left(  a\right)  }F^{\left(  a\right)  \lambda\sigma}%
)}{2\beta^{2}}}\right)  .
\end{equation}
In the Eq.s (1,2) $R$ is the Ricci Scalar, $\Lambda$ is the cosmological
constant, $K$ is the trace of the extrinsic curvature $K^{\mu\nu}$ of boundary
$\partial\mathcal{M}$ of the manifold $\mathcal{M}$, with induced metric
$\gamma_{ij}$, $\ $and $\beta$ is the Born-Infeld (BI) parameter with the
dimension of mass. Here the YM field is defined as
\begin{equation}
\mathbf{F}^{\left(  a\right)  }=\mathbf{dA}^{\left(  a\right)  }+\frac
{1}{2\sigma}C_{\left(  b\right)  \left(  c\right)  }^{\left(  a\right)
}\mathbf{A}^{\left(  b\right)  }\wedge\mathbf{A}^{\left(  c\right)  }%
\end{equation}
in which $C_{\left(  b\right)  \left(  c\right)  }^{\left(  a\right)  }$
stands for the structure constants of $\frac{n(n-1)}{2}-$ parameter Lie group
$G$ and $\sigma$ is a coupling constant. $\mathbf{A}^{\left(  a\right)  }$ are
the $SO(n)$ gauge group YM potentials. We note that the internal indices
$\{a,b,c,...\}$ do not differ whether in covariant or contravariant form.
Variation of the action with respect to the space-time metric $g_{\mu\nu}$
yields the field equations%
\begin{gather}
G_{\ \nu}^{\mu}+\frac{n\left(  n-1\right)  }{6}\Lambda g_{\ \nu}^{\mu
}=T_{\ \nu}^{\mu},\\
T_{\ \nu}^{\mu}=\frac{1}{2}g_{\ \nu}^{\mu}L\left(  \mathbf{F}\right)
+g^{\mu\alpha}\frac{2\mathbf{Tr}\left(  F_{\nu\lambda}^{\left(  a\right)
}F_{\alpha}^{\left(  a\right)  \ \lambda}\right)  }{\sqrt{1+\frac
{\mathbf{Tr}(F_{\lambda\sigma}^{\left(  a\right)  }F^{\left(  a\right)
\lambda\sigma})}{2\beta^{2}}}},
\end{gather}
where $G_{\mu\nu}$ is the Einstein tensor. Variation with respect to the gauge
potentials $\mathbf{A}^{\left(  a\right)  }$ yields the YM equations%
\begin{equation}
\mathbf{d}\left(  \frac{^{\star}\mathbf{F}^{\left(  a\right)  }}{\sqrt
{1+\frac{\mathbf{Tr}(F_{\lambda\sigma}^{\left(  a\right)  }F^{\left(
a\right)  \lambda\sigma})}{2\beta^{2}}}}\right)  +\frac{1}{\sigma}C_{\left(
b\right)  \left(  c\right)  }^{\left(  a\right)  }\frac{1}{\sqrt
{1+\frac{\mathbf{Tr}(F_{\lambda\sigma}^{\left(  a\right)  }F^{\left(
a\right)  \lambda\sigma})}{2\beta^{2}}}}\mathbf{A}^{\left(  b\right)  }%
\wedge^{\star}\mathbf{F}^{\left(  c\right)  }=0,
\end{equation}
where $^{\star}$ means duality. Our metric ansatz for $N=n+1$, is chosen as
\begin{equation}
ds^{2}=-f\left(  r\right)  dt^{2}+\frac{dr^{2}}{f\left(  r\right)  }%
+r^{2}d\Omega_{n-1}^{2},
\end{equation}
in which $f\left(  r\right)  $ is our metric function and
\begin{equation}
d\Omega_{n-1}^{2}=d\theta_{1}^{2}+\underset{i=2}{\overset{n-1}{%
{\textstyle\sum}
}}\underset{j=1}{\overset{i-1}{%
{\textstyle\prod}
}}\sin^{2}\theta_{j}\;d\theta_{i}^{2},
\end{equation}
where%
\[
0\leq\theta_{n-1}\leq2\pi,0\leq\theta_{i}\leq\pi,\text{ \ \ }1\leq i\leq n-2.
\]

\subsection{Energy momentum tensor}

In this subsection we calculate the energy momentum tensor defined by Eq. (7)
in $N\left(  =n+1\right)  -$dimensions. As we have recently introduced and
used the higher dimensional version of the Wu-Yang ansatz in EYM theory of
gravity \cite{5} we write the gauge potential one-forms as
\begin{align}
\mathbf{A}^{(a)}  &  =\frac{Q}{r^{2}}\left(  x_{i}dx_{j}-x_{j}dx_{i}\right)
,\text{ \ \ }Q=\text{charge, \ }r^{2}=\overset{n}{\underset{i=1}{\sum}}%
x_{i}^{2},\\
2  &  \leq j+1\leq i\leq n,\text{ \ and \ }1\leq a\leq n(n-1)/2,\nonumber
\end{align}
in which, by using (5), one gets the YM field two-forms satisfying the YM
equations\cite{5}. Nevertheless the energy momentum tensor defined by (7), is
found after using
\begin{gather}
\mathbf{Tr}(F_{\lambda\sigma}^{\left(  a\right)  }F^{\left(  a\right)
\lambda\sigma})=\frac{\left(  n-1\right)  \left(  n-2\right)  Q^{2}}{r^{4}%
},\text{ \ }\\
L\left(  \mathbf{F}\right)  =4\beta^{2}\left(  1-\sqrt{1+\frac{\left(
n-1\right)  \left(  n-2\right)  Q^{2}}{2\beta^{2}r^{4}}}\right)
\end{gather}
as%
\begin{align}
T_{t}^{t}  &  =T_{r}^{r}=2\beta^{2}\left(  1-\sqrt{1+\frac{\left(  n-1\right)
\left(  n-2\right)  Q^{2}}{2\beta^{2}r^{4}}}\right)  ,\\
T_{\theta_{i}}^{\theta_{i}}  &  =2\beta^{2}\left(  1-\sqrt{1+\frac{\left(
n-1\right)  \left(  n-2\right)  Q^{2}}{2\beta^{2}r^{4}}}\right)
+\frac{2\left(  n-2\right)  Q^{2}}{r^{4}\sqrt{1+\frac{\left(  n-1\right)
\left(  n-2\right)  Q^{2}}{2\beta^{2}r^{4}}}},
\end{align}
where $1\leq i\leq n-1.$ One may easily show that, in the limit of
$\beta\rightarrow0,$ the energy momentum tensor reduces to the pure gravity
\begin{equation}
T_{t}^{t}=T_{r}^{r}=T_{\theta_{i}}^{\theta_{i}}=0
\end{equation}
and once $\beta\rightarrow\infty,$ it becomes the EYM case \cite{5}%
\begin{equation}
T_{\text{ }b}^{a}=-\frac{\left(  n-1\right)  \left(  n-2\right)  Q^{2}}%
{2r^{4}}\text{diag}\left[  1,1,\kappa,\kappa,..,\kappa\right]  ,\text{ \ and
\ }\kappa=\frac{n-5}{n-1}.
\end{equation}
In the sequel we shall use this energy momentum tensor to find black hole
solutions to the EYMBI, EYMBIGB and EYMBIGBL field equations with/without
cosmological constant $\Lambda.$

\subsection{EYMBI Black hole solution in five dimensions}

In 5-dimensions, the EYMBI field equations (6) after some calculation, can be
written as%

\begin{gather}
3rf^{\prime}+6\left(  f-1\right)  +4\left(  \Lambda-\beta^{2}\right)
r^{2}+4\beta\sqrt{\beta^{2}r^{4}+3Q^{2}}=0,\\
\left[  r^{2}f^{\prime\prime}+4rf^{\prime}+2\left(  f-1\right)  +4\left(
\Lambda-\beta^{2}\right)  r^{2}\right]  \sqrt{\beta^{2}r^{4}+3Q^{2}}%
+4\beta\left(  \beta^{2}r^{4}+Q^{2}\right)  =0,
\end{gather}
which admits the following solution:%
\begin{gather}
f\left(  r\right)  =1-\frac{2M+\beta\left(  \beta-\sqrt{Q^{2}+\beta^{2}%
}\right)  }{r^{2}}-\frac{\left(  \Lambda-\beta^{2}\right)  }{3}r^{2}\\
-\frac{\beta}{3}\sqrt{\beta^{2}r^{4}+3Q^{2}}-\frac{Q^{2}}{r^{2}}\ln\left[
\frac{\left(  \beta r^{2}+\sqrt{\beta^{2}r^{4}+3Q^{2}}\right)  }{\sqrt
{4\beta^{2}+3Q^{2}}}\right]  .\nonumber
\end{gather}
This is a black hole solution and $M$ is an integration constant to be
identified as the mass of the black hole. One can show that in the limit of
$\beta\rightarrow\infty,$ $L\left(  \mathbf{F}\right)  $ and $f\left(
r\right)  $ reduce to the case of EYM as we mentioned above, i.e.,
\begin{equation}
\underset{\beta\rightarrow\infty}{\lim}L\left(  \mathbf{F}\right)
=\mathbf{Tr}(F_{\lambda\sigma}^{\left(  a\right)  }F^{\left(  a\right)
\lambda\sigma})=\frac{6Q^{2}}{r^{4}},\text{ \ \ \ }\underset{\beta
\rightarrow\infty}{\lim}f\left(  r\right)  =1-\frac{2M}{r^{2}}-\frac{\Lambda
}{3}r^{2}-\frac{2Q^{2}\ln\left(  r\right)  }{r^{2}},
\end{equation}
while in the limit of $\beta\rightarrow0$ they reduce to the pure gravity with
the cosmological constant
\begin{equation}
\underset{\beta\rightarrow0}{\lim}L\left(  \mathbf{F}\right)  =\mathbf{Tr}%
(F_{\lambda\sigma}^{\left(  a\right)  }F^{\left(  a\right)  \lambda\sigma
})=0,\text{ \ \ \ }\underset{\beta\rightarrow0}{\lim}f\left(  r\right)
=1-\frac{2M}{r^{2}}-\frac{\Lambda}{3}r^{2}.
\end{equation}
The black hole solution (20) asymptotically behaves like a de-Sitter spacetime
(Anti de-Sitter) such that%
\[
\underset{r\rightarrow\infty}{\lim}f\left(  r\right)  =1-\frac{\Lambda}%
{3}r^{2}%
\]
and for $\Lambda=0,$ it is asymptotically flat. The Born-Infeld parameter
$\beta$ modifies the radius of the horizon, as we plot in Fig. (1). In fact,
for $\beta=0$ the solution matches with the pure gravity while for
$\beta=\infty$ it gives the horizon of the EYM black hole. We notice that, BI
parameter interpolates the horizon of the corresponding black hole, between
the two extremal values of the radii of the horizons for $\beta=0$ and
$\beta=\infty.$

\subsection{EYMBI black hole solution for $N\geq5$ dimensions}

In higher dimensions $N\left(  =n+1\right)  $, the EYMBI field equations
become
\begin{gather}
\left(  n-1\right)  rg^{\prime}+\left(  n-1\right)  \left(  n-2\right)
g+4r^{2}\left(  \frac{n\left(  n-1\right)  }{12}\Lambda-\beta^{2}\right)
+4\sqrt{r^{4}\beta^{4}+\frac{\left(  n-1\right)  \left(  n-2\right)  \beta
^{2}Q^{2}}{2}}=0,\nonumber\\
\sqrt{\beta^{2}r^{4}+\frac{\left(  n-1\right)  \left(  n-2\right)  }{2}Q^{2}%
}\times\\
\left(  r^{2}g^{\prime\prime}+2\left(  n-2\right)  rg^{\prime}+\left(
n-3\right)  \left(  n-2\right)  g+4\left(  \frac{\left(  n-1\right)  \left(
n-2\right)  }{12}\Lambda-\beta^{2}\right)  r^{2}\right)  +\nonumber\\
4\beta\left(  \beta^{2}r^{4}+\frac{\left(  n-2\right)  \left(  n-3\right)
}{2}Q^{2}\right)  =0,\nonumber
\end{gather}
where $g=f\left(  r\right)  -1.$ By defining a new \ radial coordinate
$\rho=\beta r,$ and introducing $\tilde{Q}^{2}=\frac{\left(  n-1\right)
\left(  n-2\right)  }{2}\beta^{2}Q^{2},$ and $\tilde{\Lambda}=4\left(
\frac{\left(  n-1\right)  \left(  n-2\right)  }{12\beta^{2}}\Lambda-1\right)
,$ these equations can be rewritten in more convenient forms as%
\begin{gather}
\left(  n-1\right)  \rho g^{\prime}+\left(  n-1\right)  \left(  n-2\right)
g+\rho^{2}\tilde{\Lambda}+4\sqrt{\rho^{4}+\tilde{Q}^{2}}=0,\\
\sqrt{\rho^{4}+\tilde{Q}^{2}}\left(  \rho^{2}g^{\prime\prime}+2\left(
n-2\right)  \rho g^{\prime}+\left(  n-3\right)  \left(  n-2\right)
g+\tilde{\Lambda}\rho^{2}\right)  +4\left(  \rho^{4}+\frac{n-3}{n-1}\tilde
{Q}^{2}\right)  =0.
\end{gather}
These admit the general solution%
\begin{align}
f\left(  \rho\right)   &  =1+g\left(  \rho\right)  =1-\frac{\tilde{M}}%
{\rho^{n-2}}-\frac{\tilde{\Lambda}}{\left(  n-1\right)  n}\rho^{2}%
-\frac{4A\left(  \rho\right)  }{\left(  n-1\right)  \rho^{n-2}}\text{ },\\
A\left(  \rho\right)   &  =\int\sqrt{\rho^{4}+\tilde{Q}^{2}}\rho^{n-3}%
d\rho=\frac{\left\vert \tilde{Q}\right\vert }{n-2}\rho^{n-2}\text{ }_{2}%
F_{1}\left(  \frac{n-2}{4},\frac{-1}{2},\frac{n+2}{4},-\frac{\rho^{4}}%
{\tilde{Q}^{2}}\right)
\end{align}
where $\tilde{M}$ is an integration constant related to the mass of the black
hole and $_{2}F_{1}$ stands for the hypergeometric function.

\section{Field equations and the metric ansatz for EYMBIGB gravity}

The EYMBI-Gauss-Bonnet (EYMBIGB) action in $N(=n+1)-$dimensions may be written
as%
\begin{equation}
S=\frac{1}{16\pi}\int_{\mathcal{M}}d^{n+1}x\sqrt{-g}\left(  R-\frac{n\left(
n-1\right)  }{3}\Lambda+\alpha\mathcal{L}_{GB}+L\left(  \mathbf{F}\right)
\right)  +\frac{1}{8\pi}\int_{\partial\mathcal{M}}d^{n}x\sqrt{-\gamma}K\left(
\gamma\right)  ,
\end{equation}
where the terms are as before; $\alpha$ is the GB parameter (or the second
order Lovelock gravity term) and $\mathcal{L}_{GB}$ is given by
\begin{equation}
\mathcal{L}_{GB}=R_{\mu\nu\gamma\delta}R^{\mu\nu\gamma\delta}-4R_{\mu\nu
}R^{\mu\nu}+R^{2}.
\end{equation}
Variation of the new action with respect to the space-time metric $g_{\mu\nu}$
yields the field equations%
\begin{equation}
G_{\mu\nu}^{E}+\alpha G_{\mu\nu}^{GB}+\frac{n\left(  n-1\right)  }{6}\Lambda
g_{\mu\nu}=T_{\mu\nu},
\end{equation}
where
\begin{equation}
G_{\mu\nu}^{GB}=2\left(  -R_{\mu\sigma\kappa\tau}R_{\quad\nu}^{\kappa
\tau\sigma}-2R_{\mu\rho\nu\sigma}R^{\rho\sigma}-2R_{\mu\sigma}R_{\ \nu
}^{\sigma}+RR_{\mu\nu}\right)  -\frac{1}{2}\mathcal{L}_{GB}g_{\mu\nu}\text{ ,}%
\end{equation}
in which $T_{\mu\nu}$ is given in Eq. (7) and the YM field equations were
presented in Eq. (8).

\subsection{EYMBIGB black hole solution in $N=5-$dimensions}

In five dimensions, Eq. (30) leads to a set of two equations as follows%
\begin{gather}
\left(  12\alpha g-3r^{2}\right)  g^{\prime}-6rg-4r^{3}\left(  \Lambda
-\beta^{2}\right)  -4\beta r\sqrt{\beta^{2}r^{4}+3Q^{2}}=0,\\
\left[  \left(  4\alpha g-r^{2}\right)  g^{\prime\prime}+4\left(  \alpha
g^{\prime}-r\right)  g^{\prime}-2g-4r^{2}\left(  \Lambda-\beta^{2}\right)
\right]  \sqrt{\beta^{2}r^{4}+3Q^{2}}-4\beta\left(  \beta^{2}r^{4}%
+Q^{2}\right)  =0,
\end{gather}
and these equations admit an exact solution in the form of
\begin{align}
f_{\pm}\left(  r\right)   &  =1+g=1+\frac{r^{2}}{4\alpha}\left\{  1\pm\left[
1+\frac{8\alpha}{3}\left(  \Lambda+\beta^{2}\left(  \sqrt{1+\frac{3Q^{2}%
}{\beta^{2}r^{4}}}-1\right)  \right)  +\right.  \right. \\
&  \left.  \left.  \frac{16\alpha}{r^{4}}\left(  \alpha+M+\frac{1}{2}%
\beta\left(  \beta-\sqrt{Q^{2}+\beta^{2}}\right)  +\frac{Q^{2}}{2}\ln\left[
\frac{\left(  \beta r^{2}+\sqrt{\beta^{2}r^{4}+3Q^{2}}\right)  }{\sqrt
{4\beta^{2}+3Q^{2}}}\right]  \right)  \right]  ^{\frac{1}{2}}\right\}
\nonumber
\end{align}
in which $M$ is an integration constant and will be identified as the mass of
the black hole.

We notice that, this solution has the following limits:%
\begin{equation}
\underset{\beta\rightarrow\infty}{\lim}f_{\pm}\left(  r\right)  =1+\frac
{r^{2}}{4\alpha}\left\{  1\pm\left[  1+\frac{8\alpha\Lambda}{3}+\frac
{16\alpha\left(  \alpha+M\right)  }{r^{4}}+\frac{16\alpha Q^{2}\ln r}{r^{4}%
}\right]  ^{\frac{1}{2}}\right\}
\end{equation}
which is the solution of EYMGB gravity \cite{5} and $\underset{\beta
\rightarrow0}{\lim}f\left(  r\right)  $ exists if and only if $Q=0$, and one
can show that%
\begin{equation}
\underset{\beta\rightarrow0}{\lim}f_{\pm}\left(  r\right)  =1+\frac{r^{2}%
}{4\alpha}\left\{  1\pm\sqrt{1+\frac{8\alpha\Lambda}{3}+\frac{16\alpha\left(
\alpha+M\right)  }{r^{4}}}\right\}
\end{equation}
which is the case of EGB gravity. We comment that one may check $\underset
{\alpha\rightarrow0}{\lim}f_{-}\left(  r\right)  $ will produce the solution
of EYMBI gravity which was given by Eq. (20). In Fig. (2) we plot Eq. (34) for
different values of $\beta$ and fixed values for the mass, charge and
cosmological constant. We comment on this figure that, again $\beta$ provides
such a flexibility to the black hole to have any value for the radius of the
horizon between the two extremal values (i.e. the minimum value is the radius
of the horizon of the pure gravity black hole$\left(  \beta=0\right)  $, and
the maximum value corresponds with the horizon of the EYMGB black hole
$\left(  \beta=\infty\right)  $).

The positive branch of the solution is defined once $\alpha\neq0,$ and for the
positive value for $\alpha,$ the metric function $f_{+}\left(  r\right)  $ is
positive. One may find the asymptotic behavior of the metric function at large
$r,$ to show that
\begin{equation}
\lim_{r\rightarrow\infty}f_{+}\left(  r\right)  =1-\frac{\Lambda_{eff}}%
{3}r^{2}%
\end{equation}
where%
\begin{equation}
\Lambda_{eff}=-\frac{1+\sqrt{1+\frac{8\alpha}{3}\Lambda}}{4\alpha},\text{
\ \ }\alpha\neq0,\text{ \ \ }\Lambda\geq-\frac{3}{8\alpha}.
\end{equation}
This implies that $f_{+}\left(  r\right)  $ is Asymptotically-anti de Sitter
(A--AdS)-non black hole solution with an effective cosmological constant
$\Lambda_{eff}$. Finally we comment that the positive branch of the solution
with a negative value for $\alpha$ is a black hole solution which
asymptotically behaves like dS i.e.%
\begin{equation}
\lim_{r\rightarrow\infty}f_{+}\left(  r\right)  =1-\frac{\Lambda_{eff}}%
{3}r^{2}%
\end{equation}
where%
\begin{equation}
\Lambda_{eff}=\frac{1+\sqrt{1-\frac{8\left\vert \alpha\right\vert }{3}\Lambda
}}{4\left\vert \alpha\right\vert },\text{ \ \ \ }\alpha\neq0,\text{
\ \ }\Lambda\leq\frac{3}{8\left\vert \alpha\right\vert }.
\end{equation}
Such analysis for negative branch of the solution also gives same results but
$\Lambda_{eff}=-\frac{1-\sqrt{1+\frac{8\alpha}{3}\Lambda}}{4\alpha}.$ In this
case for $\Lambda=0,$ one gets $\Lambda_{eff}=0,$ which is visible from Fig. (2).

\subsection{EYMBIGB black hole solution for $N\geq5$ dimensions}

In the previous chapter we have presented a black hole solution for EYMBIGB in
5-dimensions. Our attempt in this chapter is to give a general black hole
solution to the equation (30). One can show that the general EYMBIGB equation
in $N\left(  =n+1\right)  -$dimensions can be written as%
\begin{gather}
\frac{1}{2r^{4}}\left[  \left(  r^{3}-2\alpha\left(  n-3\right)  \left(
n-2\right)  rg\right)  g^{\prime}+\left(  n-2\right)  r^{2}g-\left(
n-2\right)  \left(  n-3\right)  \left(  n-4\right)  \alpha g^{2}\right]
\left(  n-1\right)  +\\
\frac{n\left(  n-1\right)  }{6}\Lambda=2\beta^{2}\left(  1-\sqrt
{1+\frac{\left(  n-1\right)  \left(  n-2\right)  Q^{2}}{2\beta^{2}r^{4}}%
}\right)  ,\nonumber
\end{gather}
where $g=g\left(  r\right)  =f\left(  r\right)  -1.$ Again we set $\rho=\beta
r,$ $\tilde{\alpha}=\left(  n-3\right)  \left(  n-2\right)  \beta^{2}\alpha,$
$\tilde{Q}^{2}=\frac{\left(  n-1\right)  \left(  n-2\right)  }{2}$ $\beta
^{2}Q^{2}$and $\tilde{\Lambda}=4\left(  \frac{n\left(  n-1\right)  }%
{12\beta^{2}}\Lambda-1\right)  $ to get the above equation in a more
convenient form as%
\begin{equation}
4\rho^{2}\sqrt{\rho^{4}+\tilde{Q}^{2}}+\left(  \rho^{2}-2\tilde{\alpha
}g\right)  \rho\left(  n-1\right)  g^{\prime}-\tilde{\alpha}\left(
n-1\right)  \left(  n-4\right)  g^{2}+\rho^{2}\left(  n-1\right)  \left(
n-2\right)  g+\tilde{\Lambda}r^{4}=0.
\end{equation}
This equation admits the following solution%
\begin{equation}
f_{\pm}\left(  \rho\right)  =1+g\left(  \rho\right)  =1+\frac{\rho^{2}%
}{2\tilde{\alpha}}\left(  1\pm\sqrt{1+\frac{4\tilde{\alpha}\tilde{\Lambda}%
}{n\left(  n-1\right)  }+\frac{4\tilde{\alpha}\left(  \tilde{M}+4A\left(
\rho\right)  \right)  }{\left(  n-1\right)  \rho^{n}}}\right)  ,
\end{equation}
where $\tilde{M}$ is an integration constant to be identified as the mass of
the black hole and $A\left(  \rho\right)  $ is defined in Eq. (27). We comment
that, $\lim_{\alpha\rightarrow0}f_{-}\left(  \rho\right)  $ gives the
EYMBI-black hole solution given by (20) while in the case of $f_{+}\left(
\rho\right)  ,$ $\alpha$ can not be zero. In latter case one gets%
\begin{equation}
\lim_{r\rightarrow\infty}f_{+}\left(  \rho\right)  =1-\frac{\tilde{\Lambda
}_{eff}}{3}\rho^{2},\text{ \ \ }\lim_{r\rightarrow\infty}f_{+}\left(
r\right)  =1-\frac{\Lambda_{eff}}{3}r^{2},
\end{equation}
where%
\begin{equation}
\Lambda_{eff}=\beta^{2}\tilde{\Lambda}_{eff}=\left\{
\begin{tabular}
[c]{lll}%
$\frac{3\beta^{2}}{2\tilde{\alpha}}\left(  1+\sqrt{1+\frac{4\tilde{\alpha
}\left(  \tilde{\Lambda}+4\right)  }{n\left(  n-1\right)  }}\right)  ,$ &
$\tilde{\alpha}>0,$ & $\left(  \tilde{\Lambda}+4\right)  \geq-\frac{n\left(
n-1\right)  }{4\tilde{\alpha}}$\\
$-\frac{3\beta^{2}}{2\left\vert \tilde{\alpha}\right\vert }\left(
1+\sqrt{1-\frac{4\left\vert \tilde{\alpha}\right\vert \left(  \tilde{\Lambda
}+4\right)  }{n\left(  n-1\right)  }}\right)  ,$ & $\tilde{\alpha}<0,$ &
$\left(  \tilde{\Lambda}+4\right)  \leq\frac{n\left(  n-1\right)
}{4\left\vert \tilde{\alpha}\right\vert }$%
\end{tabular}
\ \ \ \ \right.
\end{equation}
which implies for $\tilde{\alpha}>0$($\tilde{\alpha}<0$), the solution is
A-dS(A--AdS) with an $\beta-$independent effective cosmological constant
$\Lambda_{eff}.$ Similar to the 5-dimensional case the negative branch of the
solution admits a $\Lambda_{eff}=\frac{3\beta^{2}}{2\tilde{\alpha}}\left(
1-\sqrt{1+\frac{4\tilde{\alpha}\left(  \tilde{\Lambda}+4\right)  }{n\left(
n-1\right)  }}\right)  ,$ with proper values for $\tilde{\Lambda},$ and
$\tilde{\alpha}.$ In this case It is also easy to show that, $\Lambda_{eff}$
is $\beta-$independent and for $\Lambda=0$($\tilde{\Lambda}=-4,$ therefore)
the effective cosmological constant vanishes.

\section{Field equations and the metric ansatz for EYMBIGB-Lovelock gravity}

In this section we consider a more general action which involves, beside the
GB term, the third order Lovelock term. The EYMBIGBL action in $N(=n+2)-$%
dimensions (we notice that in the case of EYMBIGBL $n=N-2$ and therefore it
differs from before which was chosen as $n=N-1$ ), is given by%
\begin{equation}
S=\frac{1}{16\pi}\int_{\mathcal{M}}d^{n+2}x\sqrt{-g}\left(  R-\frac{n\left(
n+1\right)  }{3}\Lambda+\alpha_{2}\mathcal{L}_{GB}+\alpha_{3}\mathcal{L}%
_{(3)}+L\left(  \mathbf{F}\right)  \right)  +\frac{1}{8\pi}\int_{\partial
\mathcal{M}}d^{n+1}x\sqrt{-\gamma}K\left(  \gamma\right)  ,
\end{equation}
where $\alpha_{2}$ and $\alpha_{3}$ are the second and third order Lovelock
parameters, and\cite{6}
\begin{align}
\mathcal{L}_{\left(  3\right)  } &  =2R^{\mu\nu\sigma\kappa}R_{\sigma
\kappa\rho\tau}R_{\quad\mu\nu}^{\rho\tau}+8R_{\quad\sigma\rho}^{\mu\nu
}R_{\quad\nu\tau}^{\sigma\kappa}R_{\quad\mu\kappa}^{\rho\tau}\nonumber\\
&  +24R^{\mu\nu\sigma\kappa}R_{\sigma\kappa\nu\rho}R_{\ \mu}^{\rho}%
+3RR^{\mu\nu\sigma\kappa}R_{\sigma\kappa\mu\nu}\nonumber\\
&  +24R^{\mu\nu\sigma\kappa}R_{\sigma\mu}R_{\kappa\nu}+16R^{\mu\nu}%
R_{\nu\sigma}R_{\ \mu}^{\sigma}\\
&  -12RR^{\mu\nu}R_{\mu\nu}+R^{3},\nonumber
\end{align}
is the third order Lovelock Lagrangian. Variation of the new action with
respect to the space-time metric $g_{\mu\nu}$ yields the field equations%
\begin{equation}
G_{\mu\nu}+\alpha_{2}G_{\mu\nu}^{GB}+\alpha_{3}G_{\mu\nu}^{\left(  3\right)
}+\frac{n\left(  n+1\right)  }{6}\Lambda g_{\mu\nu}=T_{\mu\nu},
\end{equation}
where
\begin{gather}
G_{\mu\nu}^{\left(  3\right)  }=-3\left(  4R_{\qquad}^{\tau\rho\sigma\kappa
}R_{\sigma\kappa\lambda\rho}R_{~\nu\tau\mu}^{\lambda}-8R_{\quad\lambda\sigma
}^{\tau\rho}R_{\quad\tau\mu}^{\sigma\kappa}R_{~\nu\rho\kappa}^{\lambda
}\right.  \\
+2R_{\nu}^{\ \tau\sigma\kappa}R_{\sigma\kappa\lambda\rho}R_{\quad\tau\mu
}^{\lambda\rho}-R_{\qquad}^{\tau\rho\sigma\kappa}R_{\sigma\kappa\tau\rho
}R_{\nu\mu}+8R_{\ \nu\sigma\rho}^{\tau}R_{\quad\tau\mu}^{\sigma\kappa
}R_{\ \kappa}^{\rho}\nonumber\\
+8R_{\ \nu\tau\kappa}^{\sigma}R_{\quad\sigma\mu}^{\tau\rho}R_{\ \rho}^{\kappa
}+4R_{\nu}^{\ \tau\sigma\kappa}R_{\sigma\kappa\mu\rho}R_{\ \tau}^{\rho
}-4R_{\nu}^{\ \tau\sigma\kappa}R_{\sigma\kappa\tau\rho}R_{\ \mu}^{\rho
}\nonumber\\
+4R_{\qquad}^{\tau\rho\sigma\kappa}R_{\sigma\kappa\tau\mu}R_{\nu\rho}%
+2RR_{\nu}^{\ \kappa\tau\rho}R_{\tau\rho\kappa\mu}+8R_{\ \nu\mu\rho}^{\tau
}R_{\ \sigma}^{\rho}R_{\ \tau}^{\sigma}\nonumber\\
-8R_{\ \nu\tau\rho}^{\sigma}R_{\ \sigma}^{\tau}R_{\ \mu}^{\rho}-8R_{\quad
\sigma\mu}^{\tau\rho}R_{\ \tau}^{\sigma}R_{\nu\rho}-4RR_{\ \nu\mu\rho}^{\tau
}R_{\ \tau}^{\rho}\nonumber\\
+4R_{\quad}^{\tau\rho}R_{\rho\tau}R_{\nu\mu}-8R_{\ \nu}^{\tau}R_{\tau\rho
}R_{\ \mu}^{\rho}+4RR_{\nu\rho}R_{\ \mu}^{\rho}\nonumber\\
\left.  -R^{2}R_{\nu\mu}\right)  -\frac{1}{2}\mathcal{L}_{\left(  3\right)
}g_{\mu\nu}.\nonumber
\end{gather}
The Eq.(48) after making substitutions, reads%
\begin{gather}
12\rho^{4}\sqrt{\rho^{4}+\tilde{Q}^{2}}+3n\rho\left(  \rho^{4}-2\rho^{2}%
\tilde{\alpha}_{2}g+3\tilde{\alpha}_{3}g^{2}\right)  g^{\prime}+\nonumber\\
3n\tilde{\alpha}_{3}\left(  n-5\right)  g^{3}-3n\rho^{2}\tilde{\alpha}%
_{2}\left(  n-3\right)  g^{2}+3n\rho^{4}\left(  n-1\right)  g+\tilde{\Lambda
}\rho^{6}=0,
\end{gather}
where $g=g\left(  \rho\right)  =f\left(  \rho\right)  -1,$ $\rho=\beta r,$
$\tilde{\alpha}_{2}=\beta^{2}\left(  n-1\right)  \left(  n-2\right)
\alpha_{2},$ $\tilde{\alpha}_{3}=\beta^{4}\left(  n-1\right)  \left(
n-2\right)  \left(  n-3\right)  \left(  n-4\right)  \alpha_{3},$ $\tilde
{Q}^{2}=n\left(  n-1\right)  \beta^{2}Q^{2}/2$ and $\tilde{\Lambda}%
=\frac{n\left(  n+1\right)  }{\beta^{2}}\Lambda-12.$

\subsection{7-dimensional EYMBIGBL black hole solution}

The latter equation (50) in seven dimensions which is the minimum
dimensionality of spacetime to see the effect of the third order Lovelock
gravity, by setting $n=5,$ reads
\begin{equation}
12\rho^{3}\sqrt{\rho^{4}+\tilde{Q}^{2}}+15\left(  \rho^{4}-2\rho^{2}%
\tilde{\alpha}_{2}g+3\tilde{\alpha}_{3}g^{2}\right)  g^{\prime}-30\rho
\tilde{\alpha}_{2}g^{2}+60\rho^{3}g+\tilde{\Lambda}\rho^{5}=0.
\end{equation}
This admits a solution%
\begin{equation}
f\left(  \rho\right)  =1+g\left(  \rho\right)  =1+\frac{\tilde{\alpha}_{2}%
}{3\tilde{\alpha}_{3}}\rho^{2}+\frac{\sqrt[3]{\xi}}{30\tilde{\alpha}_{3}%
}-\frac{10\left(  3\tilde{\alpha}_{3}-\tilde{\alpha}_{2}^{2}\right)  \rho^{4}%
}{3\tilde{\alpha}_{3}\sqrt[3]{\xi}},
\end{equation}
where we have used the following abbreviations%
\begin{align}
\xi &  =-4500\tilde{\alpha}_{2}\rho^{6}\left(  \tilde{\alpha}_{3}-\frac{2}%
{9}\tilde{\alpha}_{2}^{2}\right)  -150\left(  \tilde{\Lambda}\tilde{\alpha
}_{3}\rho^{6}-2\sqrt{\chi}+72\left(  A+\frac{m}{12}\right)  \tilde{\alpha}%
_{3}\right)  \tilde{\alpha}_{3},\nonumber\\
\chi &  =300\left(  \tilde{\alpha}_{3}-\frac{1}{4}\tilde{\alpha}_{2}%
^{2}\right)  \rho^{12}+15\left(  \tilde{\Lambda}\rho^{6}+6m+72A\right)
\tilde{\alpha}_{2}\rho^{6}\left(  \tilde{\alpha}_{3}-\frac{2}{9}\tilde{\alpha
}_{2}^{2}\right)  +\frac{1}{4}\tilde{\alpha}_{3}^{2}\left(  \tilde{\Lambda
}\rho^{6}+6m+72A\right)  ^{2},\nonumber\\
A  &  =\int\rho^{3}\sqrt{\rho^{4}+\tilde{Q}^{2}}d\rho=\frac{1}{6}\left(
\rho^{4}+\tilde{Q}^{2}\right)  ^{\frac{3}{2}}.
\end{align}
The metric function (52) at large values for $\rho$( and $r$ therefore) reads%
\begin{equation}
f\left(  \rho\right)  =1-\frac{\tilde{\Lambda}_{eff}}{3}\rho^{2},\text{
\ \ }f\left(  r\right)  =1-\frac{\Lambda_{eff}}{3}r^{2},
\end{equation}
where%
\begin{equation}
\Lambda_{eff}=\beta^{2}\tilde{\Lambda}_{eff}=\beta^{2}\left(  \frac{10\left(
3\tilde{\alpha}_{3}-\tilde{\alpha}_{2}^{2}\right)  }{\tilde{\alpha}_{3}%
\eta^{1/3}}-\frac{10\tilde{\alpha}_{2}+\eta^{1/3}}{10\tilde{\alpha}_{3}%
}\right)
\end{equation}
in which%
\begin{align}
\eta &  =200\left(  5\tilde{\alpha}_{2}^{3}-\tilde{\alpha}_{3}^{2}\right)
-150\tilde{\alpha}_{3}\left(  30\tilde{\alpha}_{2}+\tilde{\Lambda}%
\tilde{\alpha}_{3}-\sqrt{\chi}\right)  ,\\
\chi &  =\left(  \tilde{\Lambda}+12\right)  ^{2}\tilde{\alpha}_{3}%
^{2}+20\tilde{\alpha}_{2}\left(  \tilde{\Lambda}+12\right)  \left(
3\tilde{\alpha}_{3}-\frac{2}{3}\tilde{\alpha}_{2}^{2}\right)  +300\left(
4\tilde{\alpha}_{3}-\tilde{\alpha}_{2}^{2}\right)  .\nonumber
\end{align}
One can show that, $\Lambda_{eff}$ is $\beta-$independent and for the case of
zero cosmological constant (i.e. $\Lambda=0$ or $\tilde{\Lambda}=-12$)$,$
$\Lambda_{eff}$ \ vanishes.

As a specific choice, for technical reasons, we set $3\tilde{\alpha}%
_{3}-\tilde{\alpha}_{2}^{2}=0,$ (i.e. $\tilde{\alpha}_{3}=\tilde{\alpha}%
_{2}^{2}/3$) then this solution reduces to the simpler form
\begin{equation}
f\left(  \rho\right)  =1+\frac{\rho^{2}}{\tilde{\alpha}_{2}}\left(
1-\sqrt[3]{1+\frac{\tilde{\Lambda}\tilde{\alpha}_{2}}{30}+\frac{\tilde{\alpha
}_{2}\left(  2\left(  \rho^{4}+\tilde{Q}^{2}\right)  ^{\frac{3}{2}}+\tilde
{M}\right)  }{5\rho^{6}}}\right)  ,
\end{equation}
which is an asymptotically flat black hole solution. This solution may be
expressed as an explicit function of $\beta$%
\begin{equation}
f\left(  r\right)  =1+\frac{r^{2}}{\hat{\alpha}_{2}}\left(  1-\sqrt[3]%
{1+\left(  \Lambda-\frac{2}{5}\beta^{2}\right)  \hat{\alpha}_{2}+\frac
{3\hat{\alpha}_{2}M}{r^{6}}+\frac{2\hat{\alpha}_{2}\beta^{2}}{5}\left(
1+\frac{10Q^{2}}{\beta^{2}r^{4}}\right)  ^{\frac{3}{2}}-\frac{4\hat{\alpha
}_{2}Q^{3}\sqrt{10}}{\beta r^{6}}}\right)
\end{equation}
where $\hat{\alpha}_{2}=12\alpha_{2}.$ This expression, clearly in the two
extremal limits gives%
\begin{align}
\underset{\beta\rightarrow0}{\lim}f\left(  r\right)   &  =1+\frac{r^{2}}%
{\hat{\alpha}_{2}}\left(  1-\sqrt[3]{1+\Lambda\hat{\alpha}_{2}+\frac
{3\alpha_{2}M}{r^{6}}}\right)  ,\\
\underset{\beta\rightarrow\infty}{\lim}f\left(  r\right)   &  =1+\frac{r^{2}%
}{\hat{\alpha}_{2}}\left(  1-\sqrt[3]{1+\Lambda\hat{\alpha}_{2}+\frac
{3\hat{\alpha}_{2}M}{r^{6}}+\frac{6\hat{\alpha}_{2}Q^{2}}{r^{4}}}\right)  .
\end{align}
From (58) it is observed that asymptotically $(r\rightarrow\infty)$ we obtain
an effective cosmological constant given by $\Lambda_{eff}=\frac{3}%
{\hat{\alpha}_{2}}\left[  \left(  1+\Lambda\hat{\alpha}_{2}\right)
^{1/3}-1\right]  $ which vanishes for $\Lambda=0.$ Eq. (58) and its extremal
limits are plotted in Fig.s (3) and (4) for different values for $\Lambda$. It
is clear that for $\Lambda=0,$ $\ f\left(  r\right)  $ is Asymptotically
flat-black hole while for $\Lambda\neq0,$ $\ f\left(  r\right)  $ would be
either A-dS or A-AdS depending on the values of $\hat{\alpha}_{2}$ and
$\Lambda.$

\subsection{EYMBIGBL black hole solution for $N\left(  =n+2\right)  \geq7$
dimensions}

In higher dimensions $N\left(  =n+2\right)  \geq7,$ in general, the master
equation given by (50) admits a solution as%
\begin{equation}
f\left(  \rho\right)  =1+\frac{\tilde{\alpha}_{2}}{3\tilde{\alpha}_{3}}%
\rho^{2}+\frac{\sqrt[3]{\xi}}{6n\tilde{\alpha}_{3}\rho^{n-5}}-\frac{2\left(
3\tilde{\alpha}_{3}-\tilde{\alpha}_{2}^{2}\right)  n\rho^{n-5}}{3\tilde
{\alpha}_{3}\sqrt[3]{\xi}},
\end{equation}
where
\begin{align}
\xi &  =-\frac{36n^{2}\rho^{2\left(  n-5\right)  }}{n+1}\left\{  \tilde
{\alpha}_{2}n\left(  n+1\right)  \rho^{n+1}\left(  \tilde{\alpha}_{3}-\frac
{2}{9}\tilde{\alpha}_{2}^{2}\right)  \right.  +\\
&  \left.  \left(  \tilde{\Lambda}\tilde{\alpha}_{3}\rho^{n+1}+12\left(
n+1\right)  \left(  -\frac{\sqrt{\chi}}{36}+\left(  A+\frac{m}{12}\right)
\tilde{\alpha}_{3}\right)  \right)  \tilde{\alpha}_{3}\right\}  ,\nonumber\\
\chi &  =\frac{1}{\left(  n+1\right)  ^{2}}\left\{  \left(  -3\tilde{\alpha
}_{2}^{2}+12\tilde{\alpha}_{3}\right)  n^{2}\left(  n+1\right)  ^{2}%
\rho^{2\left(  n+1\right)  }+216n\left(  n+1\right)  \tilde{\alpha}_{2}\left(
\tilde{\alpha}_{3}-\frac{2}{9}\tilde{\alpha}_{2}^{2}\right)  \right.  \times\\
&  \left.  \left[  \frac{\tilde{\Lambda}}{12}\rho^{\left(  n+1\right)
}+\left(  A+\frac{m}{12}\right)  \left(  n+1\right)  \right]  \rho^{\left(
n+1\right)  }+1296\tilde{\alpha}_{3}^{2}\left[  \frac{\tilde{\Lambda}}{12}%
\rho^{\left(  n+1\right)  }+\left(  A+\frac{m}{12}\right)  \left(  n+1\right)
\right]  ^{2}\right\}  ,\nonumber\\
A  &  =\int\rho^{n-2}\sqrt{\rho^{4}+\tilde{Q}^{2}}d\rho=\frac{\left\vert
\tilde{Q}\right\vert }{n-1}\rho^{n-1}\text{ }_{2}F_{1}\left(  \frac{n-1}%
{4},\frac{-1}{2},\frac{n+3}{4},-\frac{\rho^{4}}{\tilde{Q}^{2}}\right)  .
\end{align}
The case of $\tilde{\alpha}_{3}=\tilde{\alpha}_{2}^{2}/3$ may be considered in
this solution and this leads us to
\begin{equation}
f\left(  \rho\right)  =1+\frac{\rho^{2}}{\tilde{\alpha}_{2}}\left(
1-\sqrt[3]{1+\frac{\tilde{\Lambda}\tilde{\alpha}_{2}}{n\left(  n+1\right)
}+\frac{\tilde{\alpha}_{2}\left(  12A+\tilde{M}\right)  }{n\rho^{n+1}}%
}\right)
\end{equation}
where $A$ is given in Eq. (27). One may use the asymptotic form of $A\left(
\rho\right)  =\rho^{n+1}/\left(  n+1\right)  $ to write
\begin{align}
\lim_{r\rightarrow\infty}f\left(  \rho\right)   &  =1-\frac{\tilde{\Lambda
}_{eff}}{3}\rho^{2},\text{ \ \ }\lim_{r\rightarrow\infty}f\left(  r\right)
=1-\frac{\Lambda_{eff}}{3}r^{2}\\
\Lambda_{eff}  &  =\beta^{2}\tilde{\Lambda}_{eff}=-\frac{3\beta^{2}}%
{\tilde{\alpha}_{2}}\left(  1-\sqrt[3]{1+\frac{\left(  \tilde{\Lambda
}+12\right)  \tilde{\alpha}_{2}}{n\left(  n+1\right)  }}\right)  ,
\end{align}
herein $\Lambda_{eff}$ is independent of $\beta$ and vanishes for $\Lambda=0$
(i.e. $\tilde{\Lambda}=-12$). Finally we comment that, for arbitrary Lovelock
parameters and $\Lambda\neq0,$ $\Lambda_{eff}$ is also defined which is
$\beta-$independent, and vanishes for $\Lambda=0.$

\section{Conclusion}

In this work we have found black hole solutions to the field equations of
EYMBI, EYMBIGB and EYMBIGBL theories of gravity. We have explicitly shown
that, these black hole solutions are the interpolated solutions between pure
gravity and gravity coupled with the YM non-Abelian gauge potentials. It is
first time that a higher dimensional non-Abelian gauge field is considered
exactly within such a context in higher dimensions. The BI parameter plays the
role of an adjustment key from the pure gravity toward EYM solutions. We
exploit this property of $\beta$ as an interpolating parameter between the two
different sets to show by numerical calculations that construction of regular
black holes become possible. Our results have been supported by some figures.
Although our treatment of the third order Lovelock parameter $\alpha_{3}$ is
constrained by the GB parameter $\alpha_{2},$ this seemed to be the only way
to compactify our expressions. Asymptotically $\left(  r\rightarrow
\infty\right)  $ once $\Lambda=0,$ in the most general case $\alpha_{2}%
\neq0\neq$ $\alpha_{3}$, by analytical calculation, it can be proved that it
gives a flat spacetime, while for $\alpha_{3}=0$ we have dS/AdS, depending on
the sign of $\alpha_{2}.$ (We notice that in the case of EYMBIGB-black hole
the positive branch of the general solution, provided us to have A-dS and
A-AdS solutions depend on the relevant parameters.) In the most general
version (i.e. EYMBIGBL) of the theory we have constructed 5-parametric black
hole solutions consisting of $(M,$ $Q,$ $\alpha_{2},$ $\alpha_{3}$ and
$\beta).$ It is our belief that with the dilatonic extension these additional
parameters will enrich string theory significantly.

\section{ Acknowledgement}

We thank to the anonymous referee for useful comments.

\section{Figure Captions}

Fig. (1): Plots of $f\left(  r\right)  $ versus $r$, for $M=1,$ $Q=1,$
$\Lambda=0$ and $\beta=0,0.1,0.5,1.0,10,1000$ and $\infty.$ The role of
$\beta$ may be interpreted as an adjustment key to get any value for the
radius of the horizon, between the extremal horizons of the corresponding pure
gravity (E)$\left(  \beta=0\right)  $ and \ EYM$\left(  \beta=\infty\right)  $
black holes.

Fig. (2): Plots of $f_{-}\left(  r\right)  $ versus $r$, for $M=1,$ $Q=1,$
$\Lambda=0$ and $\beta=0,0.1,0.5,1.0,10,50$ and $\infty.$ The role of $\beta$
may be interpreted as a regulator to get any value for the radius of the
horizon, between the horizons corresponding to pure gravity with GB
term(EGB)$\left(  \beta=0\right)  $ and EYMGB$\left(  \beta=\infty\right)  $
black holes. The smaller figure shows that by the choice of $\beta$ it is
possible to obtain black holes which are regular at $r=0$.

Fig. (3): Plots of $f\left(  r\right)  $ versus $r$, for fixed values of
$M=1,$ $Q=1,$ $\Lambda=0,$ $\alpha_{2}=1/12$ , $\alpha_{3}=1/72,$ and
$\beta=0,0.01,0.1,1,10,100,1000$, and $\infty.$ Different values of $\beta$
from $0$ to $\infty,$ corresponds to different black hole solutions between
EGBL gravity and EYMBIGBL$.$ By setting $\Lambda=0,$ the metric function
represents an A-F-black hole and therefore independent of $\beta,$ all cases
converge to a constant(=1).

Fig. (4): Plots of $f\left(  r\right)  $ versus $r$, for fixed values of
$M=1,$ $Q=1,$ $\Lambda=0.3,$ $\alpha_{2}=1/12$ , $\alpha_{3}=1/72,$ and
$\beta=0,0.01,0.1,1,10,100,1000$, and $\infty.$ Different values of $\beta$
from $0$ to $\infty,$ corresponds to different black hole solutions between
EGBL gravity and EYMBIGBL$.$ By setting $\Lambda=0.3,$ the metric function
represents an A-dS-black hole and therefore independent of $\beta,$ all cases
diverge to a -$\infty$.

\end{document}